\documentclass[%
 reprint,
 amsmath,amssymb,
 aps,
]{revtex4-1}

\usepackage{graphicx}% Include figure files
\usepackage{dcolumn}% Align table columns on decimal point
\usepackage{bm}% bold math

\usepackage{titlesec}
\usepackage{lipsum}
\usepackage{tipa}
\usepackage{parskip}
\titleformat{\section}
  {\large\bfseries}{}{0.em}{}
\titlespacing{\section}{0pt}{\parskip}{-\parskip}

\begin{document}

%Title of paper
\title{Elasticity Induced Force Reversal Between Active Spinning Particles in Dense Passive Media}

\author{J. L. Aragones, J. P. Steimel and A. Alexander-Katz \footnote{aalexand@mit.edu}}
%\homepage[]{Your web page}
%\thanks{}
%\altaffiliation{}
\affiliation{Department of Materials Science and Engineering, Massachusetts Institute of Technology, Cambridge, Massachusetts 02139, USA}

\date{\today}

\begin{abstract}
The self-organization of active particles is governed by their dynamic effective interactions. Such interactions are controlled by the medium in which such active agents reside. Here, we study the interactions between active agents in a dense non-active medium. Our system consists of actuated spinning (active) particles embedded in a dense monolayer of passive (non-active) particles. We demonstrate that the presence of the passive monolayer alters dramatically the properties of the system and results in a reversal of the forces between active spinning particles from repulsive to attractive. The origin of such reversal is due to the coupling between the active stresses and elasticity of the system. This discovery provides a new mechanism for the interaction between active agents in complex and structured media, opening up new opportunities to tune the interaction range and directionality via the mechanical properties of the medium.
\end{abstract}

\maketitle

%\section{Introduction}
Life occurs out of equilibrium. Living organisms are continuously generating and
consuming energy to achieve self-generated motion. In addition, equilibrium conditions
are barely found in nature or industrial material processing. Thus, active systems
have attracted much attention in recent years.
These systems exhibit exotic behaviors not possible under equilibrium constraints such as
emergent collective motion~\cite{Bricard:2014jq,Kumar:1ix}, 
pattern formation~\cite{Riedel:2005fz,Cates:2010hx,Farrell:2012ks}, or even phase segregation in the
absence of attractive interactions~\cite{Fily:2012hj,Redner:2013jo,Stenhammar:2013kb,Tiribocchi:2014dt}. 
The most studied active agents are those which convert some sort of energy into translational motion.
Such systems resemble how bacteria swim, and are known as self-propelled 
agents~\cite{Fily:2012hj,Redner:2013jo,Buttinoni:2013de,Wysocki:2014ky}.
Energy conversion into rotational motion is also common in nature;
important examples are the motor adenosine triphosphate synthase~\cite{Noji:1997ja},
certain cilia~\cite{Nonaka:1998vm}, the vortex array formation of sperm
cells~\cite{Riedel:2005fz}, and the dancing {\it Volvox}~\cite{Drescher:2009cy}. 
Furthermore, experiments and several numerical and theoretical studies have focused on this type of active
system in viscous media~\cite{Grzybowski:tt,Grzybowski:2002gg,Grzybowski:2001fj,Gehrig:2006iw,Lenz:2003hm,YEO:2014vv,Nguyen:2014dl,Gotze:2011cq,Llopis:2008ff,Zottl:2014fn,Goto:1hu}.
In addition to the type of activity, the medium can have a great influence on the effective interaction
between different particles, which can be particularly important for active systems.
Mixtures of active and passive particles can be used as a model system where active
particles are embedded in a complex passive system, a scenario that is prevalent in many
biological systems or processes. For example,
bacterial biofilms, where live and dead bacteria phase segregate~\cite{Chai:2011ex,Drescher:2011cv},
cell migration through tissues~\cite{Angelini:2011dm,Trepat:2009bv}, or sperm swimming through the
viscoelastic cervical mucus~\cite{Suarez:2005gc}. Although these systems are ubiquitous, very few
works have investigated these hybrid active-passive matter
systems~\cite{Kumar:1ix,McCandlish:2012ej, Ni:2014bv,Das:2014fw,Stenhammar:2015ex,Valeriani:2011ej,Angelani:2011ie}. 
McCandlish {\it et al.} reported phase segregation of active rods in the presence of passive rods,
where they point to a dynamical instability as the origin of activity-induced phase segregation;
this instability originates from the differential parallel and transversal 
diffusion coefficients coming from the anisotropy of the rods~\cite{McCandlish:2012ej}.
Ni {\it et al.} focused on the behavior of a passive particle suspension in a glassy state
doped with active agents. They observed that the presence of active particles
shift the glass transition toward higher packing fractions~\cite{Ni:2014bv} and promote
the crystallization of hard-sphere glasses~\cite{Ni:2014bva}. 
Stenhammar {\it et al.} showed that mixtures of self-propelled and passive particles phase
separate into a dense and a dilute phase~\cite{Stenhammar:2015ex}, between which the
interfacial tension is negative~\cite{2014arXiv1412.4601B}. 
However, despite all these efforts, the origin of the emergent interactions between active agents
in mixtures with passive agents remains unclear.

\begin{figure}[!htb]
\centerline{\includegraphics[clip,scale=0.4,angle=-0]{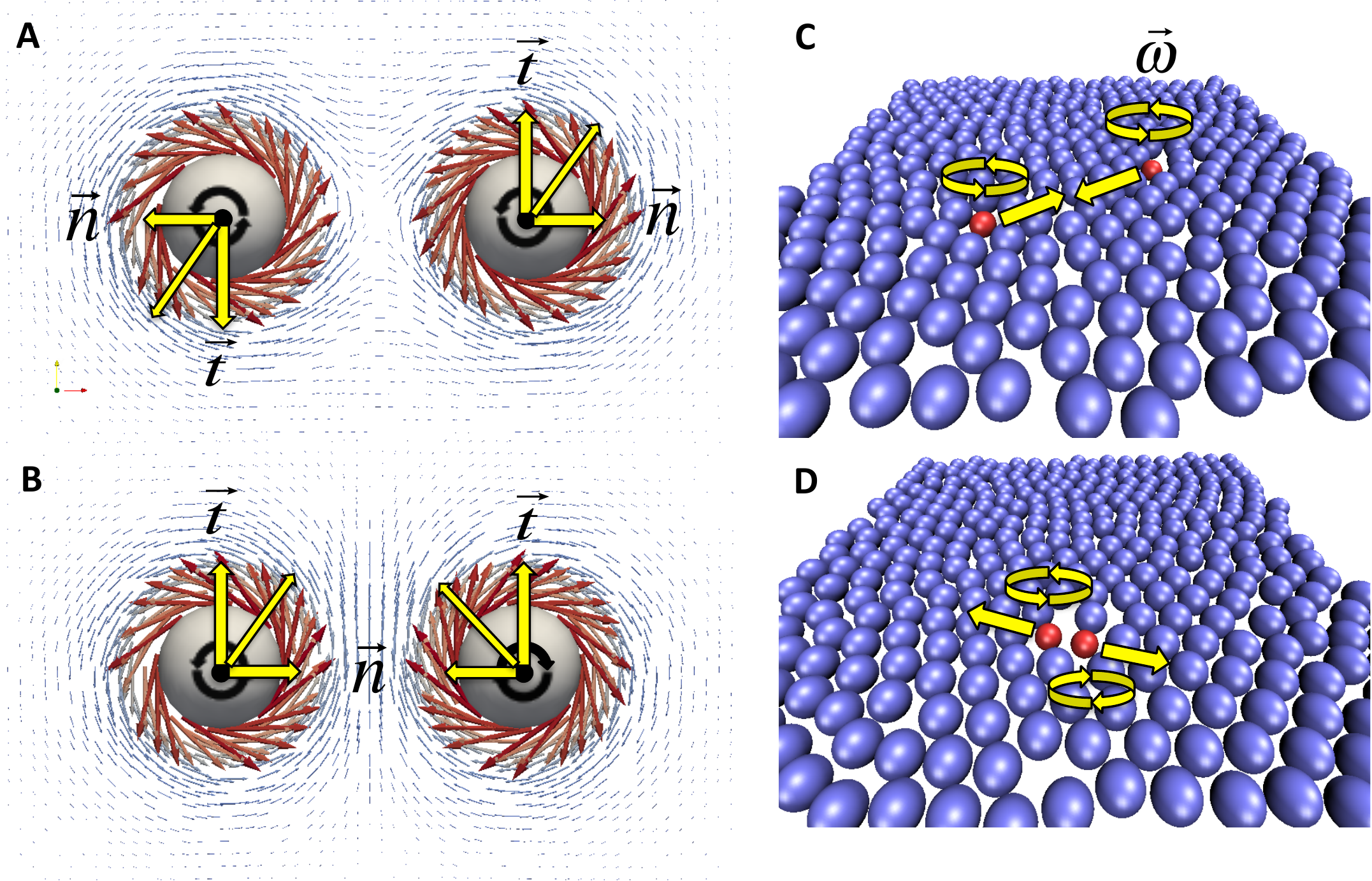}}
\caption{\textbf{(Color online) Schematic of the co- and counter-rotating spinners settled on a wall
in a viscous fluid and embedded in a dense passive monolayer.} 
A) Fluid flows generated by two co-rotating spinners in a viscous fluid medium at finite Re
result in spinner-spinner repulsion, as shown by the yellow arrows that indicate the direction
of forces exerted by the medium on the spinners. The tangential components ($\vec{t}$) come
from the fluid flows generated by the neighboring spinner, whereas the normal components ($\vec{n}$)
come from secondary flows. The resultant force generates trajectories where both spinners rotate around
their center of mass while moving apart. 
B) Fluid flows generated by two counter-rotating spinners at finite Re result in attraction.
C) Two co-rotating (red spheres) spinners rotating at frequencies $\omega$ in a dense monolayer
of passive particles (blue spheres) attract. The effective forces exerted on the spinners by the
passive medium are represented by yellow arrows.
D) Two counter-rotating spinners in a dense monolayer of passive particles repel (forces in yellow).
\label{fig1}
}
\end{figure}

To shed light on the emergent interactions that govern the self-organization of non-Brownian
active rotating particles, henceforth referred to as {\em spinners}, in systems composed of mixtures of active
and passive particles, we use both experiments and simulations. We focus on the behavior of
pairs of co-rotating and counter-rotating spinners suspended in a viscous fluid or embedded in
dense monolayers of passive particles. Importantly, we show a force reversal between spinners
as the concentration of passive particles increases above a threshold. In particular 
we observe that in a viscous fluid at small but finite Reynold numbers (Re), the fluid flows 
generated by co-rotating spinners produce a repulsion between spinners (Fig.~\ref{fig1}A),
whereas for counter-rotating spinners the resulting forces are attractive (Fig.~\ref{fig1}B).
By contrast, two co-rotating spinners in a dense passive monolayer attract each other 
(Fig.~\ref{fig1}C), whereas counter-rotating spinners repel (Fig.~\ref{fig1}D). We demonstrate that
this force reversal is induced by the change in the mechanical properties of the matrix, 
from a viscous medium, if suspended in the fluid, to a solid-like viscoelastic
medium, in the presence of passive particles. We anticipate that this mechanical attraction
between co-rotating spinners is responsible for the phase separation between active and
passive particles in macroscopic systems. 

\section{Results}
Our experimental system is composed of spherical ferromagnetic particles of diameter $\sigma$
coupled to an external rotating magnetic field of frequency $\omega$, see Fig. S1. For the experimental conditions,
the rotational frequency of the particles always coincides with the rotational frequency of the field. 
Spinners suspended in an incompressible fluid ($\nabla {\mathbf u} = 0$) of viscosity $\eta$
and density $\rho$ generate fluid flows that can be described by the Navier-Stokes equation:
\begin{eqnarray}
\label{ns}
Re\left( {\mathbf u} \cdot \nabla {\mathbf u} \right) = -\nabla p + \nabla^2 {\mathbf u} + {\mathbf f}
\end{eqnarray} 
where we have assumed no-slip boundary conditions at the particle surface,
${\mathbf u} = U + \omega \times {\mathbf r}$ being U  the translational velocity
of the particle, $\omega$ the angular velocity and ${\mathbf r}$ the vector pointing
from the center of mass of the particle to the surface of the particle. In Eq.~(\ref{ns}) we have
chosen a translating reference frame at the center of the spinner (i.e. such as the flow is steady)
and scaled the velocities and lengths by $\omega \sigma$ and $\sigma$, respectively. 
${\mathbf u}$ corresponds to the fluid velocity field, $Re$ is the Reynold
number ($Re = \omega \sigma^2 \rho / \eta$), $p$ is the pressure and ${\mathbf f}$
is the force density exerted by the particle on the fluid. Therefore, in the absence
of any other particle, and in the limit of $Re = 0$ (i.e. where the left hand terms
in Eq.~(\ref{ns}) are 0), a rotating spherical particle generates a velocity field given by~\cite{Fily:2012gq}
\begin{equation}
\label{rotating_flow}
u(r)=\frac{\tau}{8 \pi \eta r^3} \hat{z} \times {\mathbf r}
\end{equation}
where ${\mathbf r}$ is the position of the fluid from the center of the particle,
$\tau$ is the torque acting on the particle, and $\frac{\tau}{\pi \eta \sigma^3}$
corresponds to the angular rotational frequency (${\mathbf \omega}$) of the spinner,
which is constant in our system. This rotating field decays as $1/r^2$ from
the center of the spinner, as shown in Figs. S2-S4. 

\begin{figure}
\centerline{\includegraphics[clip,scale=0.35,angle=-0]{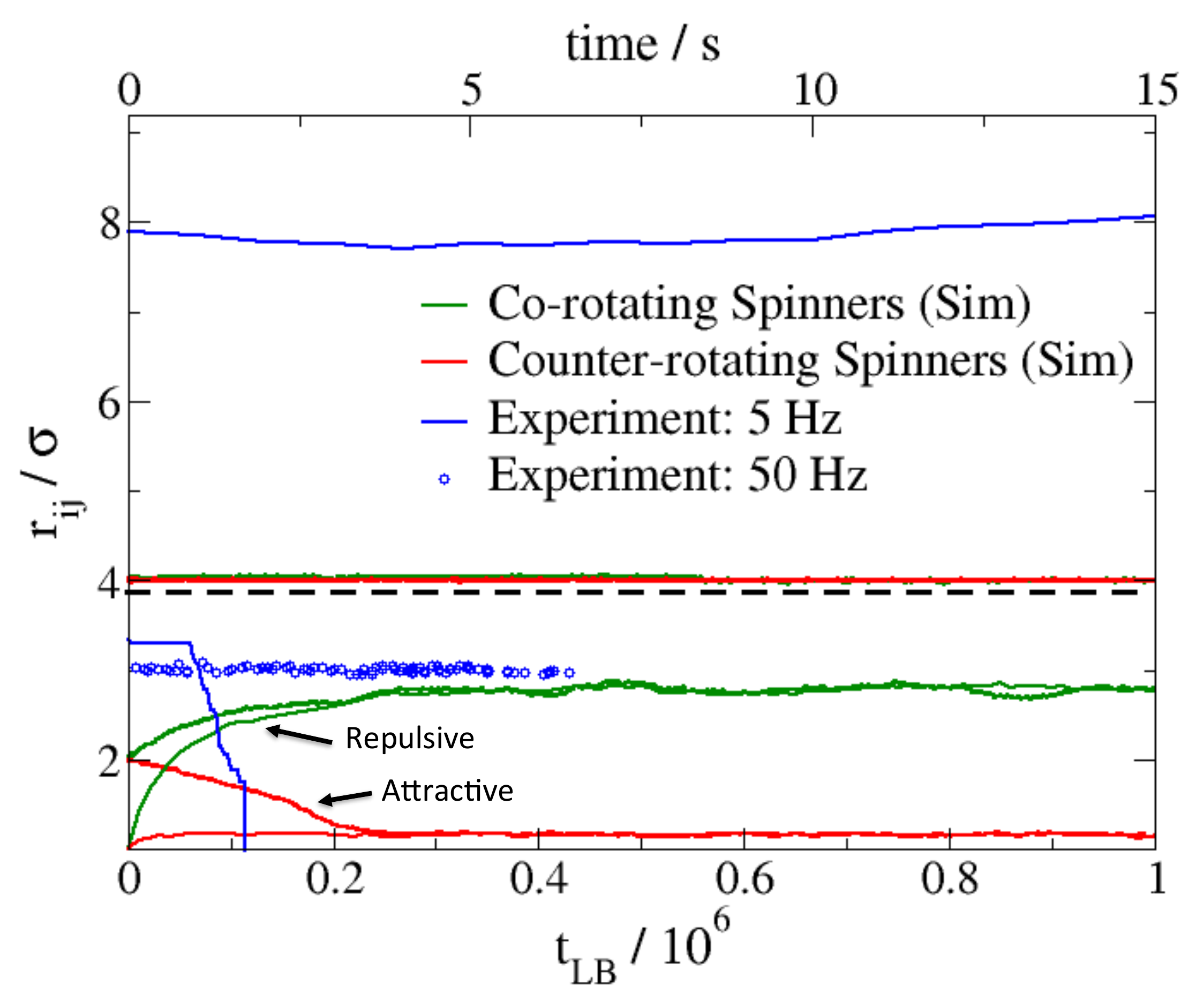}}
\caption{\textbf{(Color online) Evolution of the distance between spinners in an incompressible fluid.} 
Time evolution of the distance between two co-rotating, r$_{ij}$, in the experiments at 5~Hz
(Re  = $1.25\times10^{-3}$) (blue lines) and 50~Hz (Re = $1.25\times 10^{-2}$) (blue circles), in our simulation
model (green lines) at Re = 0.84; and between two counter-rotating spinners (red lines) using simulations at Re = 0.84. 
The dashed line at 4$\sigma$ points out the experimental interaction threshold between spinners suspended
in an incompressible fluid. The spinners are on the bottom wall of a channel of hight h = 30 $\Delta$x.
\label{fig2}}
\end{figure}

First, we study the effective interaction between two spinners in a dilute system of
co-rotating spinners in the absence of passive particles. In this system, the interaction
between spinners is controlled by the fluid flows generated by the rotation of the
spinners and the magnetic dipole-dipole attraction between them. Experimentally, we
observe that spinners initially positioned further than 4 particle diameters do not feel
either the fluid flow created by other spinners or the permanent dipole of the other spinners.
Therefore, they rotate in place without experiencing any translation, as shown in Fig.~\ref{fig2}.
By contrast, spinners closer than 4$\sigma$ attract due to the magnetic dipole-dipole interaction,
thereby forming a doublet and rotating around its center of mass. To get a deeper insight in the
behavior of the system, we also perform hybrid molecular dynamic simulations of the spinners
sedimented onto a wall within a channel of h = 30 $\Delta$x and coupled to a Lattice-Boltzmann fluid~\cite{Dunweg:2008hb}.
These simulations, which lack the dipole-dipole interaction, show that co-rotating spinners
closer than 3$\sigma$ experience a hydrodynamic repulsion, while rotating around the
center of mass of the repulsive pair. We hypothesize that in our experiments, the hydrodynamic repulsion is
hidden by the strong dipole-dipole interaction between ferromagnetic particles. To test this, we
increase the rotation frequency of the applied magnetic field up to 50~Hz. At this frequency the
hydrodynamic repulsion overcomes the dipole-dipole attraction, and spinners separate up to 
3$\sigma$, blue circles in Fig.~\ref{fig2}. The fact that the hydrodynamic repulsion increases
with Re proves the inertial nature of the interaction~\cite{Grzybowski:tt}, and it is in good agreement 
with previous observations on  millimeter-sized rotating magnetic disks adsorbed at the air-water
interface~\cite{Grzybowski:tt,Grzybowski:2002gg}.  
Although our experimental setup does not provide us with control over the direction of
rotation of individual spinners, our simulations allow us to explore the case of spinners
rotating on opposite directions. In this case, we find that two counter-rotating spinners closer than 3$\sigma$
attract each other until the separation distance between them becomes about 1.15$\sigma$ (see Fig. S5),
and simultaneously translate as a doublet in the direction orthogonal to the vector joining both 
centers~\cite{Leoni:2011gh}. The equilibrium distance between counter-rotating
spinners does not depend on the Re, but the strength of the interaction and translational velocity 
of the center of mass does (see Fig. S5), which indicates that the
observed attraction between counter-rotating spinners is also inertial in nature.

\begin{figure}
\centerline{\includegraphics[clip,scale=0.38,angle=-0]{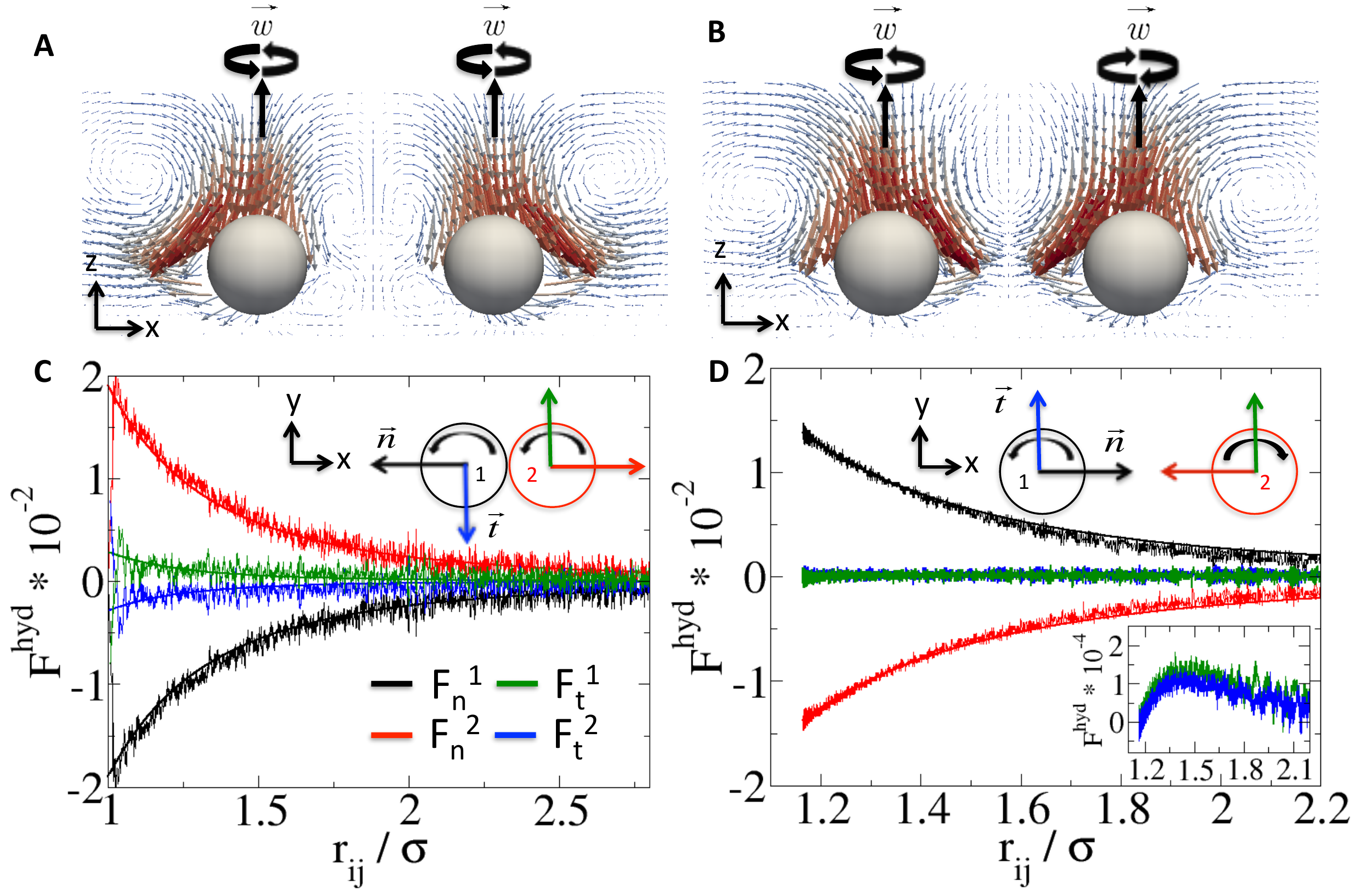}}
\caption{\textbf{(Color online) Interactions between spinners in a viscous fluid.}
Polar flow fields generated by two co-rotating spinners (A) and by two counter-rotating spinners (B)
that have sedimented onto a wall within a channel of h = 30 $\Delta x$. Hydrodynamic forces acting on
each spinner as a function of the distance between them for two co- (C) and counter-rotating
spinners (D). Solid lines are fits to the data with the function $a_0/x^3$. 
\label{fig3}}
\end{figure}

Inertial contributions to the fluid velocity field, left hand terms in Eq.~(\ref{ns}), are the origin
of the repulsion between co-rotating spinners~\cite{Climent:2007du,Grzybowski:2001fj}
and the attraction between counter-rotating spinners. At a finite Re, inertial
terms generate additional forces on the particles due to the momentum of the fluid.
These type of forces, known as lift forces, originate from the relative translation of
a rotating particle with respect to the fluid~\cite{SIRubinow:2004vx}, known as Magnus forces,
or by the translation of the rotating particle with a shear flow~\cite{Saffman:2005tf}.
Both lift forces depend on the translational velocity of the rotating particle. 
Under these conditions, the fluid velocity profile generated by a rotating sphere, Eq.~(\ref{rotating_flow}),
needs to be corrected to include these inertial terms, which generate a so-called secondary flow.
Perturbation methods have been used to calculate the secondary flow around a rotating sphere due to small
inertial effects~\cite{Bird:ws}; these studies have shown that the secondary flow produces no correction 
in the azimuthal part of the fluid velocity profile (Eq.~(\ref{rotating_flow}). However, because of the centrifugal force effect,
the fluid is pulled in toward the poles and expelled from the equator, which generates a secondary
flow on the zx-plane (see Fig. S6). The presence of a second rotating sphere breaks the symmetry 
of the secondary flow; around the equator of the spheres the fluid velocity between the spinners
decreases for co-rotating spinners and increases for counter-rotating spinners, as shown in 
Fig.~\ref{fig3}. We compute the forces exerted by the fluid on co- and counter-rotating spinner pairs
along trajectories of repulsion and attraction, as shown in Figs.~\ref{fig3}C and D, respectively.
For two co-rotating spinners, hydrodynamic forces generate a net repulsion between them, while
a hydrodynamic attraction is generated for the case of counter-rotating spinners. The fluid flows
generated by two co-rotating spinners cause the spinners to rotate around their center of mass.
This translation of the spinners generate a lift force that results in a repulsion of co-rotating
spinners~\cite{Grzybowski:2001fj,Gehrig:2006iw}. Previous studies have shown that both co-rotating
and counter-rotating spinners, repel each other as a consequence of the lift forces~\cite{Climent:2007du};
however, in the presence of a channel the hydrodynamic attraction between two counter-rotating spinners
overcome the lift force, resulting in an effective attraction~\cite{Goto:1hu}.
As discussed below, the equilibrium distance between counter-rotating spinners does not
depends on the Re number (Fig. S5); however, it does on the channel height (Fig. S7).
Thus, the confinement of the counter-rotating pair reduces the strength of the lift forces, making the
hydrodynamic attraction dominant. Similarly, the confinement of co-rotating spinners
pairs results in a shorter repulsion distance (Fig. S7) because of the reduction of the lift
force magnitude~\cite{Goto:1hu}.

\begin{figure}
\centerline{\includegraphics[clip,scale=0.4,angle=-0]{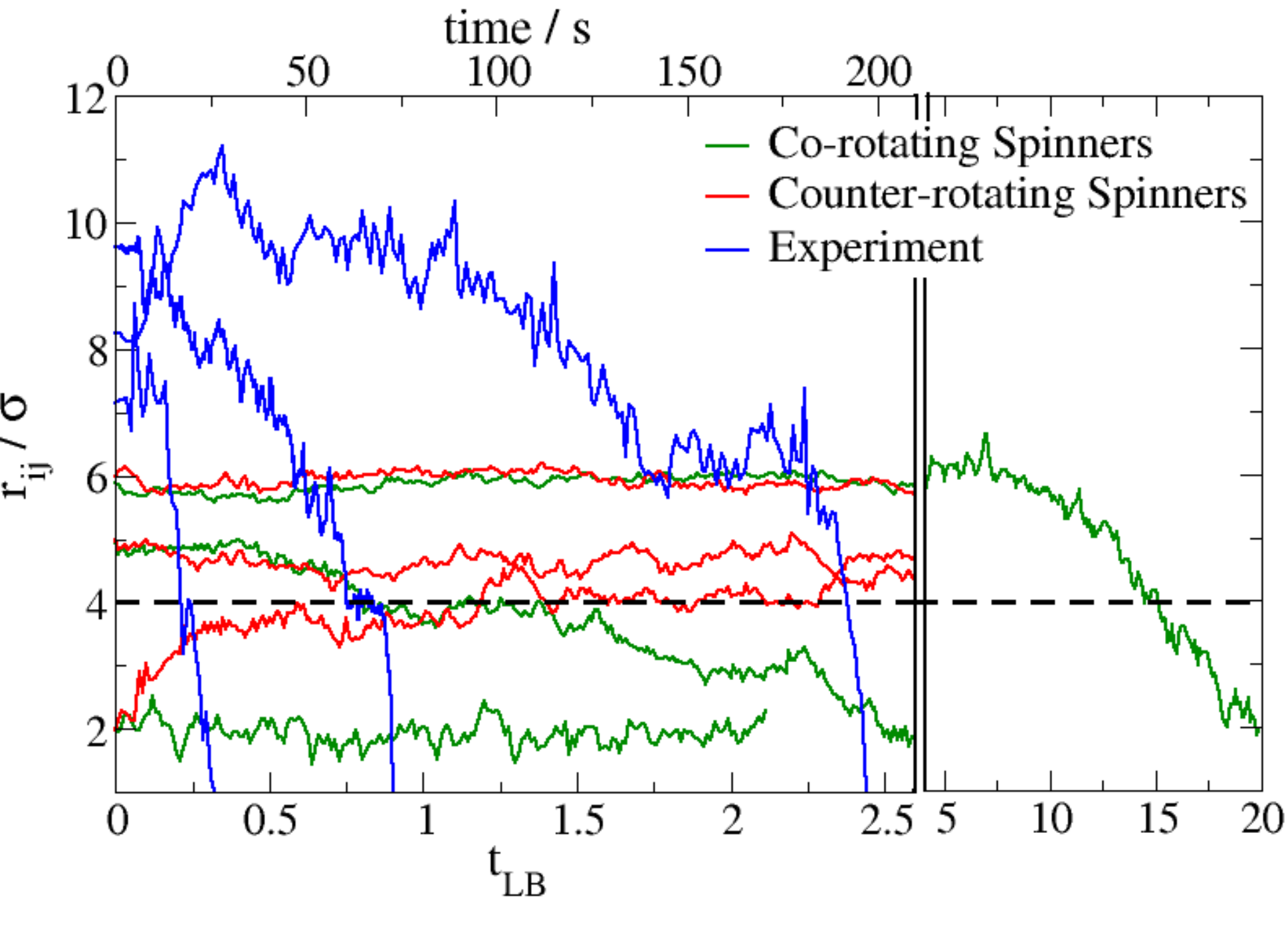}}
\caption{\textbf{(Color online) Evolution of the distance between spinners embedded in a passive particle monolayer.}
Time evolution of the distance between two co-rotating spinners in the experiments at Re = $1.25\times10^{-3}$
(blue lines) and in our simulation model at Re = 0.84 (green lines); and between two counter-rotating spinners
(red lines) at Re = 0.84 using simulations. In the simulations the monolayer area fraction is $\phi_A = 0.8$, 
whereas in the experiments is of about $\phi_A = 0.7 \pm 0.1$. The dashed line at 4$\sigma$ points out the
experimental interaction threshold between spinners suspended in an incompressible fluid.
The spinners are on the bottom wall of a channel of hight h = 30 $\Delta$x.
\label{fig4}}
\end{figure}

The behavior of spinners embedded in monolayers of passive particles is completely different.
At finite Re, the fluid flow generated by a spinner repels neighboring passive particles, and
generates the rotation of the first shell of particles around it (see Movie 1 and 2 in the SI). 
The distance of this first shell of passive particles with respect to the spinner depends 
on the area fraction of the monolayer, $\phi_A$, and the angular rotational frequency of the 
spinner, $\omega$. The effect of the rotational flow imposed by the spinner reaches 3-4 shells 
of passive particles, and the velocity of rotation of the different shells decays rapidly. 
Therefore, the spinner produces a local increase of the mobility of neighboring passive particles
and compresses the monolayer (see Fig. S8). When more than one spinner is present in the monolayer,
we observe that two co-rotating spinners attract each other; this behavior is opposite to that observed
in the absence of passive particles, as shown in both experiments and simulations in Fig.~\ref{fig4}.
The experimental trajectories of two co-rotating spinners in a monolayer with an area fraction of 
$\phi_A = 0.7 \pm 0.1$ show two well differentiated regimes: i) If the distance between the spinners
is smaller than 4$\sigma$, the slope of the trajectory is sharp; this indicates that the attraction between
spinners in this regime is governed by the strong magnetic dipole-dipole interaction. ii) At distances
larger than 4$\sigma$, the slope is small, which indicates that the attraction between spinners must
be of a different nature. By contrast, in the simulations the trajectories exhibit a single regime of slow attraction due to the lack of dipole-dipole interactions in our model. Furthermore, once the spinners
squeeze out all the passive particles initially positioned between them, they remain as a doublet at a
distance of about 2$\sigma$ for monolayers of $\phi_A = 0.8$. In our experiments we can only study
co-rotating spinners, a limitation absent in our simulations. Thus, using simulations we find that
two counter-rotating spinners repel each other at a distance of about 5$\sigma$ within a monolayer
of $\phi_A = 0.8$. Therefore, we also observe a reversal of the interaction force between two 
counter-rotating spinners with respect to the pure viscous media in the presence of a passive matrix. 

\begin{figure}
\centerline{\includegraphics[clip,scale=0.4,angle=-0]{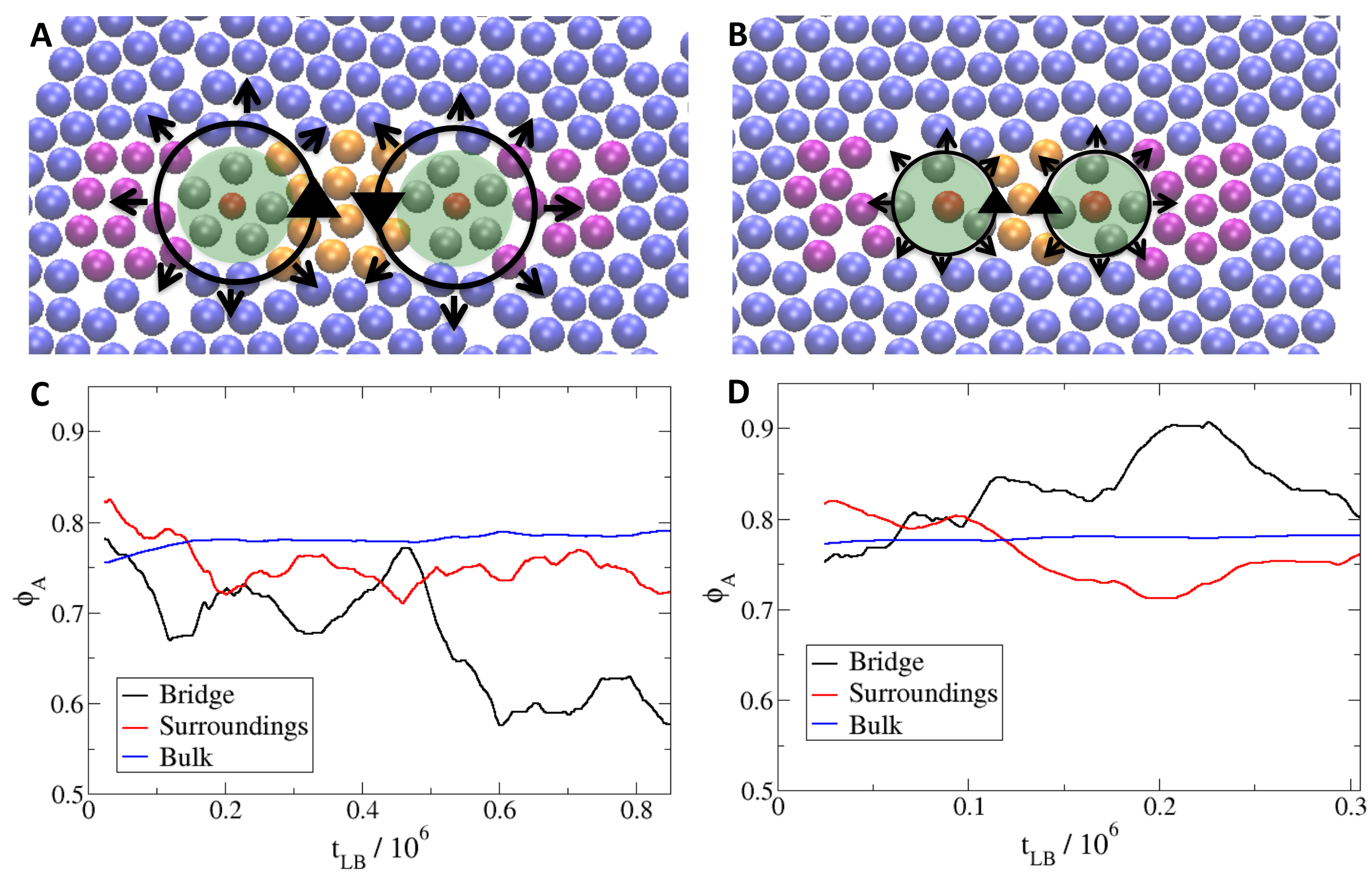}}
\caption{\textbf{(Color online) Interactions between spinners in a passive monolayer of $\mathbf{\phi_A}$ = 0.8.}
A and B) Illustration showing the forces on the system and the different regions we define
in the system: i) The corona, in green shade, which includes the spinner (red sphere) and the particles
around it (grey spheres), ii) the bridge, particles located between the spinners
(yellow spheres), iv) the surroundings, particles located besides the spinners on the opposite side
of the bridge (purple spheres), and iv) the bulk (blue spheres).
C) Time evolution of the particle area fraction of the bridge (black line), the surroundings (red line),
and the bulk (blue line) for co-rotating spinners separated by 6$\sigma$.
D) Time evolution of the particle area fraction of the bridge (black line), the surroundings (red line),
and the bulk (blue line) for counter-rotating spinners separated by 6$\sigma$.
The position of the spinners is frozen. 
\label{fig5}}
\end{figure}

To investigate the nature of the interaction between two spinners in the presence of the passive 
matrix, we define four different regions in the system and label the particles within these regions
accordingly. These four regions are: i) The first shell of particles around the spinners, named as 
corona, ii) the region between the two spinners, referred as bridge, iii) the region besides the spinners 
on the opposite side of the bridge, denoted as the surroundings, and iv) the bulk, as illustrated in Figs.~\ref{fig5}A and B.
Particles located in the corona rotate coherently around the spinners, and collide against neighboring
particles transferring their momentum. These particles rarely escape from this region and the
number of particles in the corona remains almost constant until the coronas of the two spinners 
starts to collide with each other. Therefore, we count these particles as a part of the spinner for every
calculation (green shade region in Figs.~\ref{fig5}A and B).
The bridge and the surroundings are very dynamic; particles in these regions are in continuous motion.
The stresses generated by the spinners through the corona are released in these regions.
In order to relax the stresses they need to yield. 

\begin{figure}
\centerline{\includegraphics[clip,scale=0.45,angle=-0]{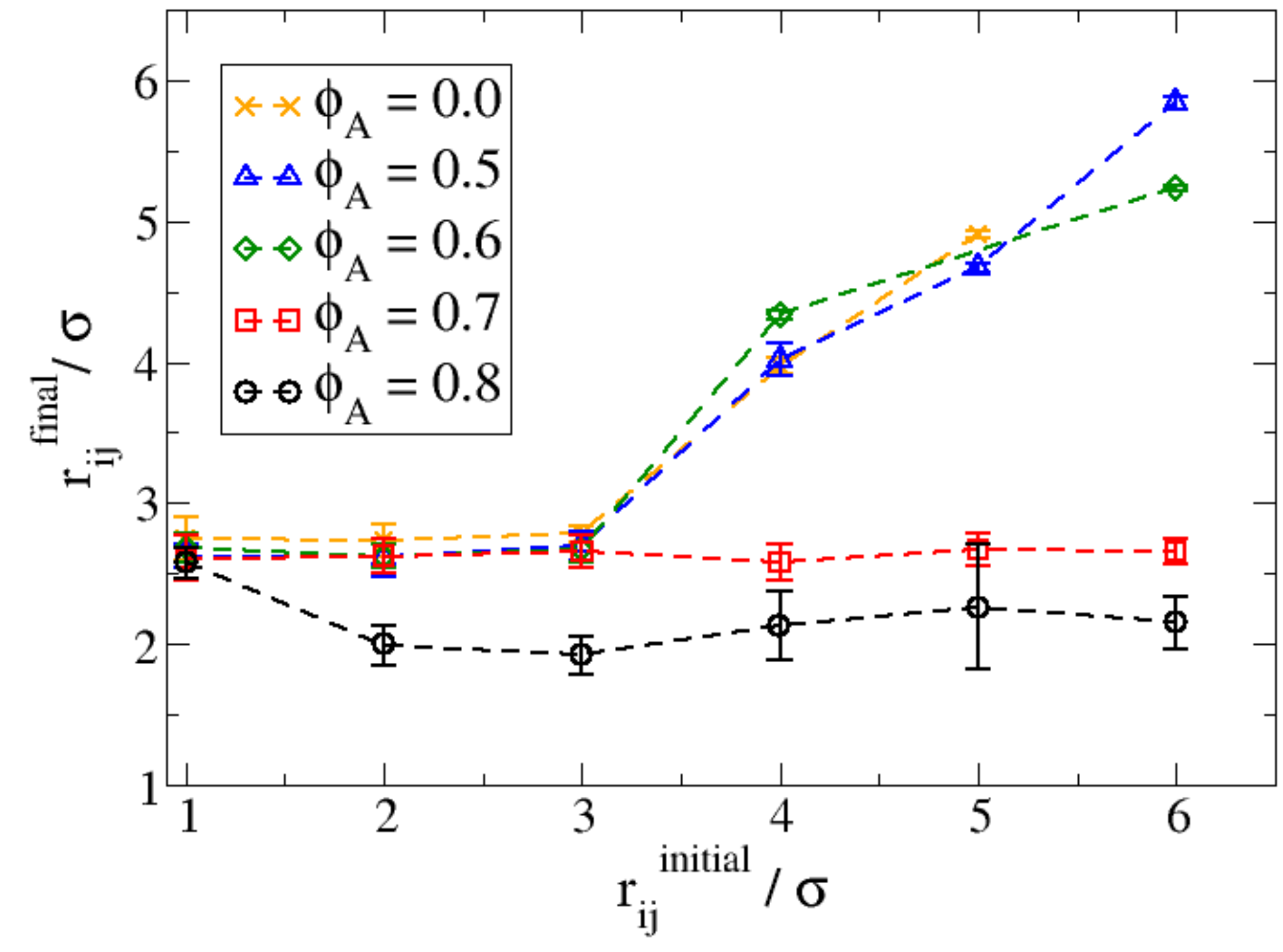}}
\caption{\textbf{(Color online) Co-rotating spinners in passive monolayers at different $\mathbf{\phi_A}$.}
Initial distance between co-rotating spinners versus the final distance after
a long simulation run at Re = 0.84 in monolayers at different particle area fractions:
$\phi = 0$ (orange crosses), $\phi = 0.5$ (blue triangles), $\phi = 0.6$ (green diamonds),
$\phi = 0.7$ (red squares) and $\phi = 0.8$ (black circles).
\label{fig6}}
\end{figure}

To study the evolution of the passive particles in the bridge, surroundings and bulk decoupled from
the spinners rearrangement, we perform numerical simulations freezing the distance between the 
spinners at different values. These constrained systems are just able to relax the stress coming 
from the activity of the spinners through the displacement of passive particles and thus, they never
reach a steady state. However, the initial time evolution of these systems allows us to study the
process of loading and yielding of the monolayer without the relaxation of the system through the
displacement of the spinners. In Figs.~\ref{fig5}C and D, the initial time evolution of the particle area
fraction in the different defined regions are presented for co-rotating and counter-rotating spinners
separated 6$\sigma$ and 4$\sigma$, respectively. We observe that for co-rotating spinners, 
the $\phi_A$ of particles in the bridge is significantly reduced as compared to the $\phi_A$ in
the surroundings and bulk, as shown in Figs.~\ref{fig5}C and S9. The compression and shear
stresses produced by the spinners in the bridge, through the corona, result in a $\phi_A$ reduction
within this region as it is constantly yielding due to shear stresses induced by the spinners.
Thus, the higher mobility of passive particles initially located in this region
allows them to migrate to less stressed regions. On the contrary, for counter-rotating spinners,
the $\phi_A$ of passive particles in the bridge is significantly increased compared to the bulk
and surroundings, as shown in Figs.~\ref{fig5}D and S9. Therefore, the mechanism
by which the passive matrix mediates the interaction between spinners is related to the type of
stresses that the spinners exert on their vicinity. Co-rotating spinners compress and shear the bridge,
as schematically illustrated in Fig.~\ref{fig5}A. To alleviate the stress, the system prefers to yield
by transporting particles from the bridge into the other regions. This occurs through avalanches
and single particle hopping, as will be shown later. Clearly, this migration reduces the density on
the bridge. This imbalance repositions the spinners closer to each other, thereby
restoring the temporal mechanical equilibrium. This process is continuously occurring, which slowly
degrades the bridge until the active particles are able to come together (see Movie 1 and 2).
During this process the monolayer is annealed, inducing the defects to migrate and concentrate
around the spinners, as depicted in Fig. S10. We discard the migration and coalescence of defects
as the driving force for the co-rotating spinners attraction by performing simulations in which the
initial configuration is a perfect hexagonal close packing (hcp) lattice. In this case, co-rotating
spinners aggregation is still observed, as shown in Fig. S11. By contrast, counter-rotating
spinners produce compression and {\em dilation} stresses in the bridge. Both spinners
move passive particles into the bridge, which increases the pressure in this region; this pushes both
spinners away, thereby resulting in a repulsion between counter-rotating spinners (see Movie 3).

\begin{figure}
\centerline{\includegraphics[clip,scale=0.35,angle=-0]{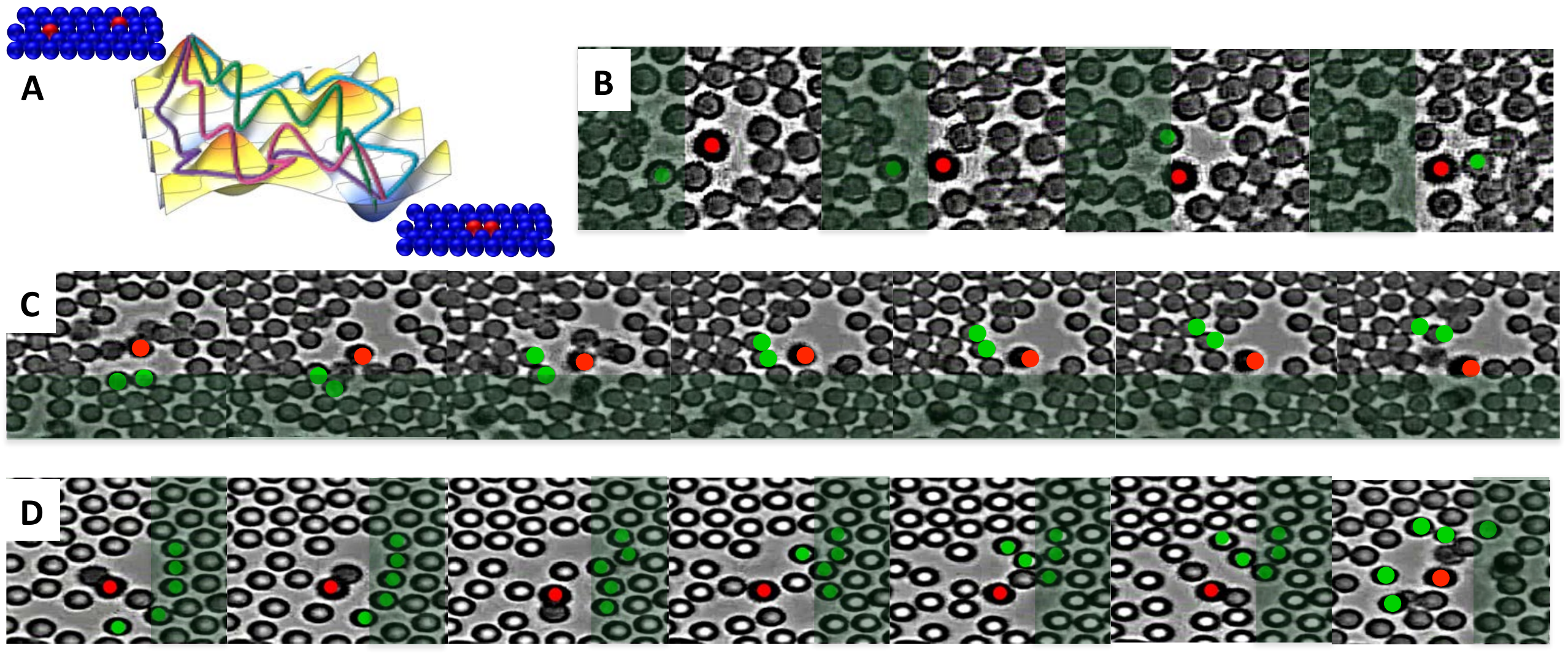}}
\caption{\textbf{(Color online) Bridge erosion mechanisms.}
A) Sketch illustrating the rough energy landscape the spinners have to move through
to form a dimer. Three different mechanism by which the passive particles are squeezed out
from the bridge by the stress imparted by the spinners: 
B) A single passive particle from the bridge jumps into the corona of one of the
spinners and is released in the surroundings or the bulk.
C) Multiple particle removal mechanism: Two or three passive particles are taken
from the bridge and moved to the bulk or surroundings.
D) Avalanche mechanism: An entire shell of particles is removed from the bridge. 
The shaded regions indicate the bridge. 
\label{fig7}}
\end{figure}

Both experiments and simulations have shown that the aggregation process of co-rotating spinners
embedded in a monolayer is governed by the elasticity of the medium and the ability of the spinners
to increase the elastic energy of the system. This can be directly confirmed by measuring the storage
and loss moduli of the system in the absence of active particles, and observing at what particle area
fraction the attractive interaction between co-rotating spinners is lost. In Fig.~\ref{fig6}, the initial distance
between two co-rotating spinners is represented against the distance reached after a long run.
For passive monolayers at $\phi_A$ = 0.8 and 0.7, spinners initially separated up to 6$\sigma$
are attracted to each other up to a distance of about 2$\sigma$ and 2.5$\sigma$, respectively.
However, for particle area fractions smaller than 0.7, passive-mediated interactions are no longer effective
and only the hydrodynamic repulsion is observed. According to the mechanical properties of the passive
monolayer, the elastic response of the system  dominates at long times for monolayers of $\phi_A$ = 0.8
and 0.7, as depicted in Figs. S13A and B; for the latter, elastic and viscous responses are almost
equivalent at long times. On the contrary, for area fractions of 0.6 and 0.5 the viscous response dominates
for the entire frequency range, Figs. S13C and D. This demonstrates that an elastic response of the
passive media is a necessary condition for co-rotating spinners to interact. Furthermore, this elasticity-mediated
attraction between active spinning particles is inherently stochastic due to the discrete nature of the passive media
(i.e. granular system). Thus, the initial configuration of the system determines the length of interaction and the time
required for spinners to aggregate, as shown by the different experimental trajectories in Fig.~\ref{fig4}.
Hence, one can imagine a rough and dynamic energy landscape in which the spinners must traverse to form
a dimer, and where multiple paths exist when moving from state to state, as schematically illustrated in
Fig.~\ref{fig7}A. Erosion of the bridge, which happens when passive particles are removed from it, occurs
through three main mechanisms: i) Single particle removal from the bridge to the corona and then to the 
bulk, as shown in Fig.~\ref{fig7}B, ii) multiple particle removal, where two or three particles in the bridge,
due to the shear stresses, are moved into the bulk or intermittently moved to the corona and then ejected
into the bulk, as seen in Fig.~\ref{fig7}C, and iii) avalanches where entire lines of particles in the bridge are
pushed into the bulk by the shear exerted by the spinners, as depicted in Fig.~\ref{fig7}D.
We use the name avalanche to invoke the instantaneous and dramatic nature of the particle removal process. 
Whereas the bridge erosion produced by the two first mechanisms is slow and depends on stochastic collisions,
the removal of particles in avalanche happens almost instantaneously. In fact, it has been previously shown that
the devitrification of hard-sphere glasses is mediated by large rearrangements of particles,
so-called avalanches~\cite{2014PNAS..111...75S}. This exciting new interaction opens up new 
possibilities to study the mechanism of elasticity-mediated interaction
between active particles in these systems governed by a glassy dynamics.   

\section{Discussion}
In summary, the forces exerted by an incompressible fluid at small but finite Re on a pair of non-Brownian
active rotating particles depends on the relative sense of rotation of each particle,
resulting repulsive for co-rotating spinners and attractive for counter-rotating
spinners when confined in a channel. The presence of a dense passive matrix modify
the mechanical properties of the system from a viscous media to a viscoelastic
material. In this latter case, the interaction between spinners becomes controlled by
elastic effects, which act in the opposite direction than inertial effects~\cite{Nadal:2014cp,Pak:2012gp,Avino:2010fu}.
Hence, the switch between inertial and elastic stresses derives in a reversal of the interaction
between spinners, resulting in an attraction of co-rotating spinners and a repulsion of counter-rotating spinners.
In fact, the structure of the passive dense medium can not be treated
at the mean field level. For example, assuming a pure viscous scenario where the passive matrix would be an
homogeneous continuum with higher viscosity than the system in the absence of passive particles, the
stress would be dissipated by the viscous media, and one would just observe the repulsion between
co-rotating spinners, but the strength of the secondary flows would be smaller due to the viscosity
increment ($Re = \omega \sigma^2 \rho / \eta$). Therefore, the change of the mechanical properties
of the matrix from a viscous material to a solid-like material is the responsible for the force reversal.    
Furthermore, the interaction between spinners in dense monolayers of passive particles is of stochastic nature;
it depends on the configuration of the passive monolayer. Thus, it is the instantaneous configuration
of the monolayer determines the strength and range of the interaction, and the dynamics of attraction
is intimately related to the timescale for the rearrangement of the monolayer. This can be better understood
by looking at the oscillation between periods of well-defined distances and periods of fast attraction along
the trajectory between two co-rotating spinners, Figs.~\ref{fig4} and S11. The level of stress put into the bridge
by the spinners must reach a configuration dependent threshold value, and when this level is reached the
system yields by removing entire groups of particles from the bridge, as depicted in Fig.~\ref{fig7}. Then,
the spinners approach each other and start stress-loading the bridge again. Therefore, this effective interaction
mediated by the passive medium cannot be seen as a position dependent interaction potential (U(r$_{ij}$)).
Interestingly, the dynamic trajectories of the distance between the spinners initially positioned at different
distances show an almost linear regime of attraction, as it can be seen in Figs.~\ref{fig4} and S11.
This means that the spinners approach each other on average at a constant speed. Assuming a
stokes' scenario, $F = 6 \pi \mu \sigma/2 U$, for the translation of the spinners through the monolayer
along their attractive trajectory, the strength would be a constant and independent on the distance between them.
Moreover, this elasticity-mediated interaction is of a very long range. For example, in our
simulations we observe that spinners separated up to 6 particle diameters still interact, whereas
the hydrodynamic interaction reach only 3 particle diameters. Remarkably, our experimental
measurements show that the interaction threshold shifts even at longer distances, and spinners
separated by up to 17$\sigma$ attract each other~\cite{Josh}, while the dipole-dipole interaction
reach 4$\sigma$ (see Figs.~\ref{fig2} and \ref{fig4}). Our results resemble other elastic media such
as lipid membranes, where elasticity-mediated forces between transmembrane embedded proteins
show logarithmic decays~\cite{Muller:2005ho,Deserno:2009iz}. However, the origin of the stresses
in our system are different. 

In conclusion, we have shown that the interaction between active spinning particles depends
on the properties of the medium and the dynamics of the active particles.
Therefore, this cooperative interaction between the mechanical and dynamical
properties of the system offers a variety of possibilities to tune this type of 
interaction between active agents. 
Remarkably, we have also observed that the spinners produce an annealing of the
passive matrix structure (Fig. S10), in agreement with previous simulation results
in hybrid passive-active systems but with a different type of active particles~\cite{Ni:2014bv}.
We anticipate that this mechanical attractive force between co-rotating
spinners is responsible for the phase separation observed in systems with
higher concentrations of spinners. Moreover, in ternary hybrid active-passive systems,
composed of mixtures of spinners rotating either clockwise or counter clockwise, we
observe phase segregation in three different phases: one composed of the passive
particles, other of co-rotating spinners, and the last one of counter-rotating 
spinners~\cite{Nguyen:2014dl,jlaragones_preparation}.  
In principle, this elasticity-induced interaction between active spinning particles is
general for other active agents such as self-propelled particles. 
In fact, preliminary results, both experimental and using simulation models, 
show that elasticity-mediated interaction between active particles also occurs for
hybrid systems composed by passive and self-propelled particles.
Therefore, this study opens up new and exiting routes to control the range and
direction of the interaction between active units in passive and structured environments.
This interaction between active spinning particles in passive matrixes is different from the
emergent interactions observed between passive objects within active fluids. Those effective
interactions mediated by active matter between passive objects depend on the mobility of
the passive objects~\cite{Ni:2015cj,Angelani:2011iea} and their shape~\cite{Harder:2014bw}.
Therefore, it would be very interesting to investigate these effects in the opposite scenario
between active particles in passive matrixes. Even more, this type of interaction could play
an important role in overcoming diffusive limitations that active biological molecules encounter
if interacting within dense viscoelastic materials such as the highly viscoelastic nucleus of the
cell~\cite{Guilak:2000ey} or cells in extracellular polysaccharide matrix secreted by biofilm
forming bacteria~\cite{Wilking:2011kw}.  

\section{Model and methods}
\linespread{0.2}\footnotesize
To study the behavior of active rotating particles in pure viscous and in dense passive environments,
we have designed a hybrid active-passive system composed by ferromagnetic particles (the active
units or spinners) and polystyrene particles (the passive units).
To measure pairwise interactions between spinners we mix an extremely dilute solution of active
ferromagnetic particles with a solution containing passive polystyrene particles. Both particles
have a diameter of 5~$\mu$m. We first study the case of pure spinners, in the absence of any
passive particle. We then study the interactions between spinners in a monolayer of passive 
particles at a particle area fraction of $\phi_A = 0.7 \pm 0.1$. These solutions, dilute spinners
and dense mixtures, are placed between a cover slip and a slide, and placed in our 
experimental setup, which is composed by four magnetic coils (Fig. S1); this allows us to generate
a rotating magnetic field that is parallel to the substrate and rotates around the z-axis
(see SI for more details). The frequency of rotation used in this study was 5~Hz.
Upon actuation of the external rotating magnetic field the magnetic dipole moment of the
spinner couples to the applied field, and the particle spins in place around the z-axis.
\\[2pt]
To gain a more detailed insight into the non-equilibrium nature of this system we carry
out numerical simulations using hybrid molecular dynamic simulations of the colloidal particles
coupled to a Lattice-Boltzmann fluid~\cite{Dunweg:2008hb}. The simulation box is discretized
in three-dimensional grids with resolution $N_x \times N_y \times N_z = 214 \times 214 \times 30$
bounded in the z direction by no-slip walls and periodic boundary conditions in the
x and y directions. The Lattice-Boltzmann (LB) fluid is described by the fluctuating
Lattice-Boltzmann equation~\cite{Dunweg:2007km}, which properly describes the 
dissipative and fluctuating hydrodynamic interactions. We implement the discrete
19-velocity model (D3Q19). The LB fluid parameters are density $\rho = 1$, kinematic
viscosity $\nu = 1/6$, and temperature $k_BT = 2 \cdot 10^{-5}$. 
For simplicity, we set the grid spacing $\Delta x$ and the LB time step $\Delta t$ equal to unity.
Interactions between the LB fluid and the particles are described by the bounce-back 
rule~\cite{Ladd:1994wb}, and enforcing no-slip boundary conditions at the surface of the particles.
Specifically, we implement the ALD method~\cite{Ding:2003wl}, where particles are treated as real solid objects. 
In our simulation model, colloidal particles are considered as hard-spheres~\cite{Jover:2012jy}
of diameter $\sigma$ = 12$\Delta$x; thus, we are just considering excluded volume interactions.
The spinner activity is generated by imposing an external torque,
$\tau$, about the z-axis. To form the monolayer we also include a gravity force F$_G$ = 0.005.
Since we are interested in the hydrodynamic interactions that occur between spinners, and not
their magnetic interaction, we do not include dipole-dipole interactions in our simulation model
to more clearly delineate the origin of the effective interactions.
\linespread{1}\normalsize
\section{Acknowledgements}
\linespread{0.2}\footnotesize
This work was supported by the U.S. Department of Energy, Office of Basic Energy Sciences,
Division of Materials Science and Engineering under award No. \#ER46919 (simulations and analysis),
the Chang Family (reagents, equipment and experiments). We are grateful to Jorn Dunkel, Ken Kamrin,
and Laura R. Arriaga for insightful discussions.
\linespread{1}\normalsize
\section{Author contributions}
\linespread{0.2}\footnotesize
JPS performed the experiments. JLA carried out the numerical simulations. JLA and JPS analyzed the results.
AAK proposed and supervised the study. JLA and AAK wrote the paper.
\linespread{1}\normalsize
\section{Additional information}
\linespread{0.2}\footnotesize
{\bf Supplementary Information} is available in the online version of the paper.
\\[2pt]
{\bf Competing financial interests:} The authors declare no competing financial interests.

%\bibliographystyle{./apsrev}
%\bibliography{./biblio}
%%%%%%%%%%%%%%%%%%%%%%%%%%%%%%%%%%%%%%%%%%%%%%%%%%%%%%%%
%
%%%%%%%%%%%%%%%%%%%%%%%%%%%%%%%%%%%%%%%%%%%%%%%%%%%%%%%%

\end{document}